# THE EMERGENCE OF POLITICAL DISCOURSE ON DIGITAL NETWORKS: THE CASE OF THE OCCUPY MOVEMENT

Jorge Fábrega

Universidad Adolfo Ibañez
Diagonal Las Torres 2640
Peñalolen, Santiago, Chile
e-mail: jorge.fabrega@uai.cl

Javier Sajuria

University College London
29/30 Tavistock Square
London, WC1H 9QU, United Kingdom
e-mail: j.sajuria@ucl.ac.uk

**ABSTRACT**

How does political discourse spread in digital networks? Can we empirically test if certain conceptual frames of social movements have a correlate on their online discussion networks? Through an analysis of the Twitter data from the Occupy movement, this paper describes the formation of political discourse over time. Building on a previous set of concepts - derived from theoretical discussions about the movement and its roots - we analyse the data to observe when those concepts start to appear within the networks, who are those Twitter users responsible for them, and what are the patterns through which those concepts spread. Preliminary evidence shows that, although there are some signs of opportunistic behaviour among activists, most of them are central nodes from the onset of the network, and shape the discussions across time. These central activists do not only start the conversations around given frames, but also sustain over time and become key members of the network. From here, we aim to provide a thorough account of the "travel" of political discourse, and the correlate of online conversational networks with theoretical accounts of the movement.

**INTRODUCTION**

The emergence of political discourse is generally a mediated experience. Either through mainstream media or elites, different discourses arouse and – sometimes – influence public opinion (Zaller 1992). Since social movements are essentially contentious against elites (Tarrow 2008), they have developed different "repertoires of action" (Tilly & Tarrow 2007) to overcome these barriers. For years, the use of illegal, violent or non-violent actions has been sufficient to raise media awareness and reach broader audiences. However, the formation of the discourse had been usually restricted by geographical and temporal constraints. Moreover, elites within the movements have been historically in charge of this process.

The advent of new technologies, particularly Twitter, allows for a new type of conversation between the members of the movements, mainstream media, and the general public (Boyd et al. 2010). In turn, these conversations propose new questions about the generation of political discourse through online interactions. This paper is an attempt to provide an answer to these questions. We use the Twitter datasets from the Occupy Research project (www.occupyresearch.net) to provide a detailed account of how political discourse is created and spreads through Twitter conversations – mainly through mentions and retweets. From there, we aim to explain how certain concepts can travel through the network and who are those Twitter users responsible for that. The formation of clusters, and the overall diversity of networks are relevant indicators to observe

The paper starts with a revision of the literature on political discourse formation. Then, it examines the Occupy movement as a research option, providing a discussion on how different political frames influenced the movement, and how they were essential throughout the months it took place. Subsequently, we provide a brief description of the datasets used for the paper, alongside the methods used to analyse them. Finally, a discussion of the results is provided, with a clear emphasis on the potential benefits of using network theory and methods to understand discourse formation.

## THE FORMATION OF POLITICAL DISCOURSE

The emergence of collective views about political topics has always been an interesting puzzle. It requires a combination of political knowledge about the issue (Mansfield & Sisson 2004; Zaller 1992), media framing (Fujiwara 2005; Gerring 1999; Schnell, Karen Callaghan 2001), and shared spaces for conversation (Boyd et al. 2010; Hampton et al. 2010).

When it comes to discourse, Berg and Lune point out that *"the interesting aspect of (...) discourse is not merely what is said, or which words are used, but the social construction and apprehension of meanings thus created through this discourse"* (2011, p.364). On that regard, political discourse refers the actual meaning of the political discussions that people hold, both privately and publicly.

Common knowledge argues that most people do not have high levels of political knowledge (Ansolabehere 2005). Hence, most discussions take place among informed elites, who have access to exclusive sources of information, and have time to learn it and use it. However, the emergence of new technologies is, allegedly, reducing the entry barriers to political information. Online political outlets expand the options for accessing political information. However, they do not necessarily increase the interest of people for getting informed (Stanley & Weare 2004; Weaver Lariscy et al. 2011). As a result of this high level of availability of political information, we might not see people acquiring more political knowledge, but having increasing difficulties distinguishing "the signal from the noise". Information and communication technologies (ICTs) then, might not be the definite solution to increase political knowledge among people.

The media plays a key role in shaping public discourse, either by framing issues in a certain way, or by expanding the views of some elites (Graber 2003; Iyengar 1987; Iyengar 1994). In the case of the recent social movements, the use of ICTs has proven to be fundamental for activists to reach mainstream media outlets (Sajuria 2013; Theocharis 2012). In that way, the actions – and omissions – of the media, shape public discourse. However, as mentioned above, the process of connecting political actors with journalists requires a message established *a priori*, which has been already defined by those leading the movement – the elites.

It is in the conversation between those related to the movement where political discourse gets created. A traditional process of hierarchical decision-making requires elites to take up people's opinions, and deliberate in closed instances. However, the use of social media is changing this process in an interesting way. Regardless of the intentions of elites to keep things exclusive, platforms such as Twitter allow any user to comment, discuss, and converse within a public environment. As boyd et al. explain,

> *"(B)ecause Twitter's structure disperses conversation throughout a network of interconnected actors rather than constraining conversations within bounded spaces or groups, many people may talk about a particular topic at once, such that others have a sense of being surrounded by a conversation, despite perhaps not being an active contributor"* (2010, p.1)

Hence, it is not only elites, or even activists who shape political discourse on Twitter. People who do not want to participate in the conversations might get exposed to them, and eventually participate.

The contribution of this paper refers to the fact that social media bring new questions and possibilities for discourse formation. The relevance of the information networks in framing political issues deserves a closer look. How do certain topics reach broader audiences, while others do not? Who are responsible for putting those topics into the main conversations? How do those processes evolve over time? This is a broader researching agenda to which this paper aims to contribute with.

## THE FRAMES BEHIND THE OCCUPY MOVEMENT

The start of the Occupy movement relates to two specific – and subsequent – events. On September 16, 2011, the Canadian online magazine Adbusters posted a call to occupy Wall Street, as a way of protest against the financial and political system. The next day, over 1,000 protesters gathered in Wall Street, and after clashing with the police, decided to occupy Zuccotti Park, a privately owned space. Within days, hundreds of public spaces were occupied around the US, and in other countries of the world (Anon n.d.)

Nevertheless, the movement was not spontaneous. After the events from the Arab Spring in early 2011, followed by the Indignados protests in Spain, the occupation of public spaces - such as Tahrir Square, Puerta del Sol or Plaza Catalunya - became an effective form of contention. The combination of a local - public - occupation, and the discussions sustained over the Internet fostered the emergence of broader frames, able to gather a larger number of people.

In terms of discourse, the Occupy movement relates closely to the Indignados protests in Spain. Both processes were framed as a struggle against inequality and abuses (Cabal 2011; Castells 2012). In fact, some scholars have suggested that the Occupy movement is the result of 30 years of class-struggle in the US (Chomsky 2012).

One of the most salient concepts amongst the Occupy protesters is the idea of "#wearethe99percent". With that, they reflect the feeling that there is a small minority of people, mostly bankers, who did not suffer the consequences of economic crises. Moreover, even if they might be deemed responsible for the crises, they usually get more money as a compensation for losing their jobs (Milkan et al. 2013). It can be said that the sense of unfairness was one of the initial drivers of the movement.

But the inequality was not the only distinctive frame of the movement. Castells (2012) argues that there are several other elements that need to be considered when analysing this, and other examples of recent social movements (such as in Egypt and Spain). Based on his work and others (Caren & Gaby 2011; Juris 2012; Occupy Research n.d.), we have established three main frames that, in our opinion, inform the discourse around the Occupy movement.

First, as already explained, the activists conversed around the topics of inequality and abuse. The frame used was "#wearethe99percent", which in turn served as a conceptual boundary between "us" (those who suffer the abuses, the victims of inequality), and "them" (those who are responsible for the economic debacle, and do not suffer any consequence from it). This frame is pre-existent to the actual Occupy movement, and some (Castells 2012) would argue that it is the initial drive for the occupations.

The second frame present in the movement refers to the importance of occupying public spaces. This is not only the material occupation of a park or a square, but also the creation of a horizontal – and seemingly democratic – space for public conversations. This frame also relates with ideas of deliberation, public spaces and horizontal democracy. In summary, the idea is that the occupation of physical and virtual spaces gives place to an ideal way of democratic discussion, where everyone is equal and has the same rights. Finally, within these frames is possible to find discussions about elites and the conception of "leaderless revolutions" (Gerbaudo 2012; Ross 2012).

The third frame is based on the central idea behind Castells' (2012) work: Outrage and Hope. The emergence of these movements is an answer to a combination of events, where communication networks play a significant role. First, economic and political crises create outrage. Then, people with access to communication networks – such as Twitter – find others who feel the same. These networks of outrage become networks of hope when – through online conversations – people realize that they are not alone, and that there are options for organization and action. The final event is action. The key of this frame is to understand when, and how, outrage becomes hope.

With these three frames, we have produced a research design intended to capture them from Twitter conversations, using social network analysis techniques.

## METHODOLOGY

### Data Sources

Twitter operates as a microblogging platform, where users post messages, each one no longer than 140 characters. Unlike other social network sites, Twitter creates directed connections. That is, that a user can follow the tweets of another user without needing the permission of that user. Moreover, the fact that user A is following user B does not ensure that user B will interact with A. Hence, relationships based on follower/following status might not always be an optimal tool to construct networks. Instead, we use interactions between users as a way to create the edges between different nodes. Interactions can take the form of a mention (when a user writes another username in their messages), a retweet (when a tweet from another user is replicated verbatim, usually with a "RT" at the beginning of it), or a "via"(when a user quotes the message from another user.

The datasets used in this paper come from the Occupy Research project (www.occupyresearch.net), a crowdsourcing enterprise to study the Occupy movements around the world. In particular, the Twitter datasets were collected by R-Shief (www.r-shief.org) using the Twitter Streaming API during a period comprised between September and December 2011. The period was selected as it corresponds to the onset of the movement, when most discourses ought to be created. From all the datasets available, we used the one containing all the messages using the "official" hashtag of the movement: #ows.

The next step was the creation of the networks. First, using keywords, a dataset for each frame was created. For the frame related to inequality, the keyword selected was "#99percent", which covers different variations of the hashtag "#wearethe99percent". The frame related to the occupation of public spaces used the term "public space". Finally, the frame related to the transformation of the networks from outrage to action, used the term "#hope" as a proxy. We acknowledge that the selection of the terms using a deductive approach carries out some limitations. The messages might contain terms that are more useful to observe the required relations, but the results obtained and the theory used to select them allows us to be confident of their validity. Nevertheless, this is an ongoing research agenda, and, therefore, further analyses will be required in the future to establish this proposition.

Second, in terms of time intervals, we decided to split the datasets using 3-day periods. This decision was made taking into consideration the way in which messages and relations are constructed on Twitter. The idea was to have a period of time large enough to capture relevant interactions among Twitter users, but short enough to allow us to observe the diffusion of each topic-centred conversation over time. Each 3-day period became a static network itself, which was, in term, analysed as part of a longitudinal series. Also, each new incomer into the conversation was identified according with the 3-days period in which she wrote her first tweets on the topic. Table 1 summarizes the data collected.

*Table 1: Descriptive statistics of the networks*

|  | 99 percent | Public Space | Hope |
|---|---|---|---|
| Total Tweets | 25189 | 4470 | 155132 |
| Retweets/Tweets | 16081 | 3197 | 89684 |
| Tweets with mentions/Tweets | 3464 | 464 | 24283 |
| Unique twitter users | 14564 | 3621 | 44864 |

As it is show in Figure 1, all frames follow similar patterns in terms of frequencies of tweets. There are two main modes for each topic during days in which the movement was particularly intense; although, the frame related with inequality and abuse shows a smoothed curve.

*Figure 1: Distribution of tweets per 3-days periods per each network*

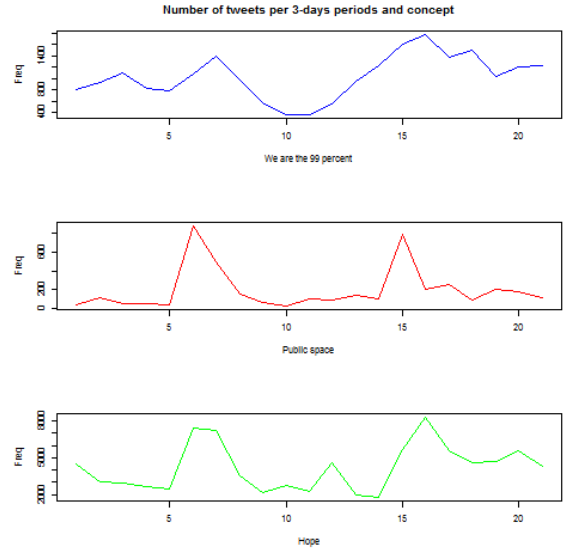

### Research Hypothesis

The exercise of understanding how many given conversation around a topic is constructed requires, mainly, a descriptive effort. In the case of the networks formed by Twitter discussions, the exercise is twofold. On the one hand, it is necessary to describe the network structure of the population. On the other, we need to provide a temporal account of how the network evolved to reach its final form.

Since the networks are formed through interactions between users, on any given topic, there will be "activists" whenever we find users who are interested in making some topics more salient. If that is the case, we should observe emergent and growing clusters of activists (by homophily) around the topic. Alternatively, the absence of significant and increasing clustering in the network might indicate that the topic attracts different individuals along time (commentators rather than activists). When activists are present, they can be remarkably efficient in starting and spreading a given message or they might follow an opportunistic behaviour, taking concepts that are already present in the conversation - but not necessarily salient - and bringing them to the centre of the discussion. In both cases, those agents will behave as connectors impacting the dissemination of opinion and consequently, we might be able to observe that role via a measure of betweenness centrality. Based on these two options, we can start describing the research hypotheses.

**H₀:** Public discourse emerges from groups of activists who have an initial goal of making an issue salient, and keep talking about it over time. If this hypothesis is correct, we should observe small and identifiable groups of activists writing and spreading their message persistently over time. Their success is then measured against their ability to connect different groups who talk about the same issue.

In terms of the measurements required to test this hypothesis, the conversational networks should be centralized around these activists from the beginning. They would be the central nodes of the network from its onset, and they will be efficient in turning the topic into a trend. In terms of the level of clustering of the network, we should not observe major variations over time, once the size of the network is taken into account.

**H₁:** Activists follow an opportunistic behaviour, taking topics that are already being discussed in the network, and activate them later. This means that the network would initially present high levels of dispersion, where there is a significant amount of "noise". Several topics would be discussed at the same time, but no one would be more relevant than the others, until the activists – mainly users that were not active in the network before – arrive and propel the issues.

In this case, we should observe a network where there are disperse, yet active nodes, with low levels of clustering. After a given period of time, the activists who decide to take over the respective frame connect these nodes.

**H₂:** There is a third possibility. It might be the case that the groups discussing about a topic change over time. That means that the frames remain constant and relevant, but the users talking about them change. In this case, we would expect a highly dispersed network with groups of users that quickly appear and disappear. This responds to the idea of cohorts and waves of participants who join the conversation at different points in time. Each wave has its own behaviour, but they all discuss within the same frames. Table 2 summarizes the heuristic behind the hypothesis

*Table 2: Expected output consistent with each hypothesis*

| Hypothesis | Clustering within cohorts | Betweenness centrality (BC) |
|---|---|---|
| Ho | High | First incomers have higher BC |
| H1 | High | Cohorts with high BC change across periods |
| H2 | Low | BC is homogeneous across cohorts and periods |

In essence, the structure of networks can be described as points within two different extremes. On the one hand, we can find one big community, where every member is connected to each other. The other option is a network with a large number of clusters, and a low – or inexistent – number of connections between them. In the case of the latter, it is safe to assume that the transmission of the topic over time did not take place through contagion. Moreover, in that case, it might be important to observe some of the other elements for the formation of discourse, such as media framing or political knowledge.

But it is when we find a large, densely connected network, when we need to start looking at its development over time. Through analysing the retweets and mentions, we can observe how conversations evolve. Depending on the speed of the process, we can assert if we are in the presence of a communicational cascade (Fabrega & Paredes 2013).

In the next section we present preliminary results. And, in order to fulfil with author guidelines we will focus the attention on the frame related with inequality.

**Analyses and results**

The main activity on Twitter associated with OWS movement extended by almost four months. However, as it is shown in figure 2, discussions on inequality and the "#wearethe99percent" concepts were adding new participants at decreasing rates. By the end of second month of the movement, most of the people who were sending new messages on inequality and related issues had records of previous participation on the topic. Consequently, it is along those two first months (or twenty periods of three days) when the discourse is framed.

*Figure 2: New incomers into the frame on inequality along time*

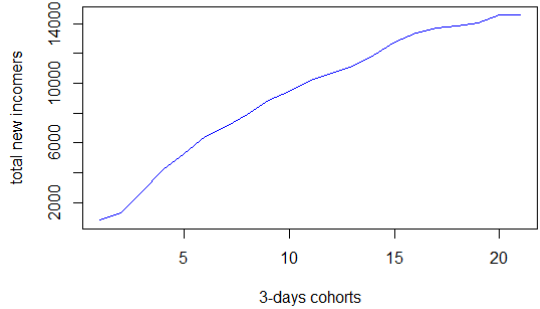

At the global level, the activity around inequality in the OWS movement shows increasing average clusterisation during the first two month (see figure 3). In our view, that increasing clusterisation was related with a process of consolidation of digital neighbourhoods around the topic. Visualizations of the social graph (Figure 4) showed us that by the end of the second month there were clear structures of giant components at the centre surrounded by a number of minor clusters on the peripheries.

*Figure 3: Average cluster coefficient by cohort in the "#wearethe99percent" frame*

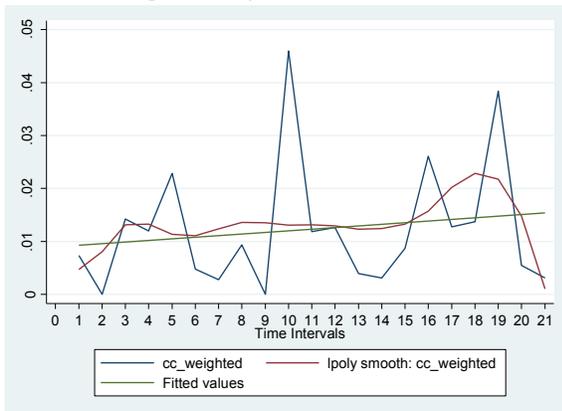

Given the concave form of the pattern of new incomers and the increasing clusterization of subsequent cohorts, the network of interactions that we have observed is apparently inconsistent with the third hypothesis presented above. It does seem to be the case that the topic of inequality is one in which conversations are casual and irregular. On the contrary, there are traces of discourse formation in the way in which interaction have evolved.

*Figure 5: Visualisations using OpenOrd algorithm of the "#wearethe99percent" network at periods 1 (1,111 tweets), 10 (973 tweets) and 21 (1,299 tweets):*

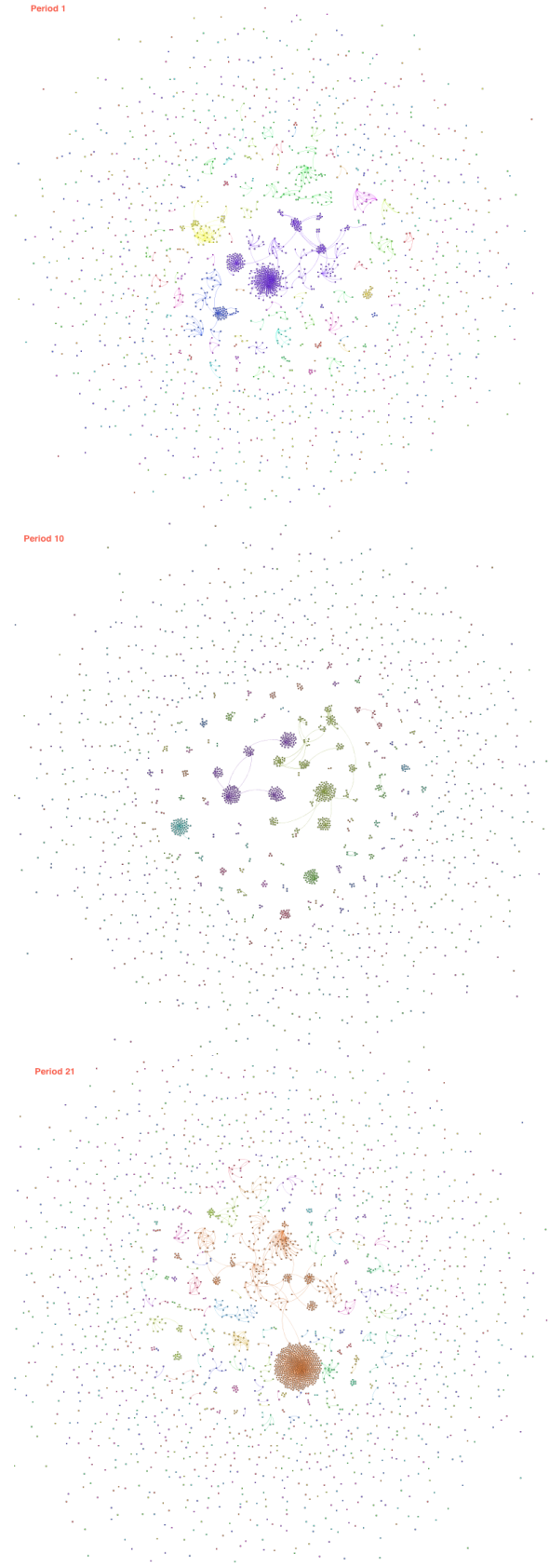

In fact, according with our analysis, early incomers into the topic of inequality become connectors in greater proportions than later participants during the period of discourse formation. Figure 5 summarizes this point. Each cell represents the average betweenness centrality of each cohorts (labeled as 1, 2,…, and 21, respectively) in a given period (labeled as bc01, bc02,…, and bc21, respectively). Blue (red) cells represent deviations to higher (lower) centrality values to the mean.

Two remarkable patterns emerge from the figure. First, the main cohort with blue cells (higher values) during the first ten periods (the first month) was cohort 1. Second, other cohorts occupy central positions for short period of time (blue cells) during the second month of the movement. These results seems to be more consistent with both hypothesis Ho and H1, but not with hypothesis H2. By the beginning of the second month, more than ten thousand users have already tweeted or retweeted on the topic using concepts like "we are the 99 percent". Consequently, it can be argued that most of the discourse formation was already in place when new groups can capture attention and centrality on the issue. Nevertheless, further analysis is required for a better understanding of the relative explanation power of both hypotheses.

*Figure 5: Average betweenness centrality by cohort during the period of discourse formation*

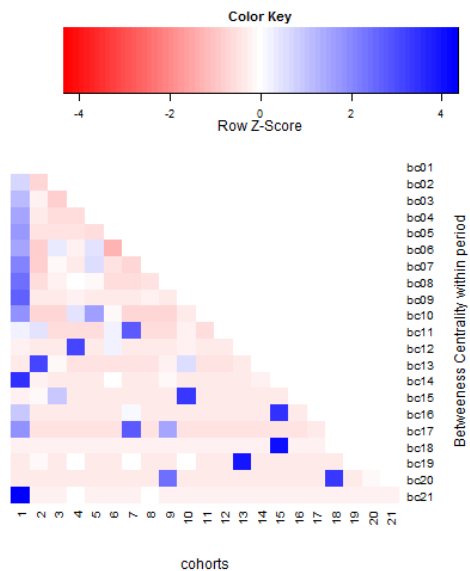

## Conclusions

The process of discourse formation in a digital era apparently does not seem to be fully decentralised nor as spontaneous as internet enthusiast would like to claim. It might be the case that groups and individuals with particular interest can take advantage of the massiveness and openness of digital media to frame the public opinion. Data from the dynamic of Twitter interaction around the Occupy Wall Street seems to be consistent with that interpretation. We think that similar patterns would be observable in other type of social phenomena on Twitter. At least, it can be said that early bird participants during the Occupy Wall Street movement got (as a group) an advantage to frame the discourse around concepts that later became key for the social movement.